\documentstyle[manuscript,aps,pra,tighten,graphicx]{revtex}
\begin{document}
\def\bea{\begin{eqnarray}} \def\eea{\end{eqnarray}}
\def\be{\begin{equation}} \def\ee{\end{equation}}
\def\rra{\right\rangle} \def\lla{\left\langle} \def\ra{\rightarrow}
\def\kvp{\bbox{k}'} \def\kv{\bbox{k}} \def\qv{\bbox{q}}
\def\ann{a_{nn}} \def\de{\Delta} \def\dmu{\delta\mu}
\def\eps{\epsilon} \def\deps{\delta e} \def\al{\alpha}
\draft

\title{Pairing in asymmetric two-component fermion matter}

\author{J. Mur-Petit\footnote{Corresponding author. Address: Dept.
    ECM, Facultat de F\'\i sica, Avda. Diagonal 647, E-08028 Barcelona
    (Spain). Phone: (+34)3 4021185. Fax: (+34)3 4021198. E-mail:
    jordim@ecm.ub.es}, A. Polls, and H.-J. Schulze}
\address{Departament d'Estructura i Constituents de la Mat\`eria,
  Universitat de Barcelona,\\
  Av. Diagonal 647, E-08028 Barcelona, Spain}
\maketitle

\begin{abstract}
  We analyze the possibilities of pairing between two different
  fermion species
  in asymmetric matter at low density.  While the direct interaction
  allows pairing only for very small asymmetries, the pairing mediated
  by polarization effects is always possible, with a pronounced
  maximum at finite asymmetry.  We present analytical results up to
  second order in the low-density parameter $k_F a$.
\end{abstract}

\pacs{PACS:
 74.20.Fg,  
 03.75.Fi,  
 24.10.Cn   
 }


The recent experimental
achievement of trapping fermionic alkali atoms at very low density and
temperature \cite{exp} demands theoretical estimates of the
characteristic size of the pairing gaps $\de$ that might be observable
in those systems.
The canonical case of an attractive $s$-wave interaction leads
(including polarization effects) to the well-known low-density result
\cite{gmb,pet,spp} \be {\de \over\mu} = {1 \over (4e)^{1/3}} {8 \over
  {e}^2} \exp\!\left[{\pi\over 2k_F a}\right] \:,
\label{e:ds}
\ee where $\mu=k_F^2/2m$ is the chemical potential and $a<0$ the
$s$-wave scattering length.  However, under certain circumstances
direct $s$-wave pairing is not possible: in the case of repulsion,
$a>0$, or, e.g., in spin-polarized one-component Fermi systems.

A similar, particular system has recently been advocated for
experimental study.  It is a spin-polarized alkali gas composed of two
different hyperfine levels of $^6$Li \cite{stoof}.  In this specific
environment an (attractive) $s$-wave interaction exists only between
the atoms at different levels, whereas atoms at like levels can only
interact via $p$-waves.  This report is dedicated to the quantitative
study of the principal possibilities of pairing in such a system, and
in particular of the dependence of pairing on the asymmetry of the
system, which evidently is an important experimental parameter that
influences strongly the magnitude of the observable gaps.

More generally, we will assume a fermionic system composed of two
distinct species 1 and 2 (carrying definite spin orientations) of
equal mass $m$
and with densities $\rho_1$ and $\rho_2$, or equivalently total
density $\rho=\rho_1+\rho_2$ and asymmetry
$\al=(\rho_1-\rho_2)/(\rho_1+\rho_2)$.  We also introduce the notation
$k_i \equiv k_F^{(i)}$, $(i=1,2)$ for the Fermi momenta
$k_i=(6\pi^2\rho_i)^{1/3}$, and $\mu_i=k_i^2 /2m$ for the two chemical
potentials at low density.

For the sake of presentation we will for the moment consider an
idealized system without direct interaction between like particles 1-1
and 2-2, whereas 1 and 2 are interacting via a
potential $V$ with a
$s$-wave scattering length $a$.  We are only interested in the
situation at very low density, $k_i|a| \ll 1$, where the pairing
properties are completely determined by the scattering length, or,
equivalently, the low-momentum $s$-wave $T$-matrix $T_0 = 4\pi a/m$.
We will now analyze the pairing gaps generated by this interaction.


In the case of attraction, $a<0$, and for very small asymmetries
clearly by far the dominant process is the pairing generated by the
direct $s$-wave interaction between different species, see
Fig.~\ref{f:dia}(a).  The BCS theory generalized to asymmetric matter
\cite{stoof,umb,akhi} yields the basic coupled equations for the
determination of the gap function $\de_k$, total density $\rho$, and
density difference $\delta\rho$, \bea \de_{k'} &=& - \sum_{k} V_{kk'}
{ \left[ 1-f(E_k^-)-f(E_k^+) \right] \over 2E_{k}} \de_{k} \:,
\\
\rho = \rho_1 + \rho_2 &=& \sum_{k} \Big[ 1 - {\eps_k\over E_k} \left[
  1 - f(E_k^-) - f(E_k^+) \right] \Big] \:,
\label{e:sr}
\\
\delta\rho = \rho_1 - \rho_2 &=& \sum_{k} \left[ f(E_k^-) - f(E_k^+)
\right] \:,
\label{e:dr}
\eea with the Fermi function $f(E) = \left[ 1 + \exp(\beta E)
\right]^{-1}$ and \be E_k^\pm = E_k \pm \dmu \ , \ E_k =
\sqrt{\epsilon_k^2 + \de_k^2} \ , \ \eps_k = e_k - \mu \ , \ e_k =
k^2\!/2m \:.
\label{e:ek}
\ee The chemical potentials of the species 1 and 2 are $\mu_{1,2} =
\mu\pm\dmu$ and $V$ is the bare potential acting between them.
Throughout this report we will only determine gaps
at zero temperature, where one has $f(E) = \theta(-E)$ and therefore
\bea \left. 1 - f(E_k^-) - f(E_k^+) \right. &=& \theta(E_k^-) \:,
\\
\left. f(E_k^-) - f(E_k^+) \right. &=& \theta(-E_k^-) \:, \eea i.e.,
the unpaired particles are concentrated in the energy interval
$[\mu-\deps ,\mu+\deps]$, $\deps = \sqrt{{\dmu}^2-{\de}^2}$, which
does not contribute to the pairing interaction.  This situation is
illustrated in Fig.~\ref{f:n}, that sketches the BCS momentum
distributions of the species 1 and 2, according to Eqs.~(\ref{e:sr})
and (\ref{e:dr}), for a positive asymmetry.  The excess of particles 1
is located in the interval around $\mu$, Pauli-blocking the gap
equation.  This leads to a rapid decrease of the resulting gap when
increasing the size $2\deps$ of the interval, i.e., the asymmetry.

In the weak-coupling case, $\de \ll \deps \ll \mu$, which is adequate
in the low-density limit, the momentum distributions of the two
species are very sharp and one obtains from Eqs.~(\ref{e:sr}) and
(\ref{e:dr}): \be \alpha =
{\delta\rho \over \rho} \approx {3\over2}{\deps \over\mu} \ll 1 \:,
\ee i.e., the asymmetry $\alpha$ is directly proportional to $\deps$.
Analyzing also the gap equation in the context of a weak-coupling
approximation, one obtains for the dependence of the gap on the
parameter $\deps$
\cite{stoof,umb}, \be \dmu + \deps = {\rm const.} = \de_0 \quad
\Leftrightarrow \quad \de^2 = \de_0^2 - 2\de_0\deps
\:, \ee where $\de_0$ is the gap in symmetric matter of the same total
density.
Consequently the gap as a function of asymmetry is: \be {\de \over
  \de_0} = \sqrt{ 1 - {4\mu \over 3\de_0}\alpha } \:.
\label{e:sgap}
\ee The gap vanishes at $\alpha_{\rm max} = 3\de_0/4\mu$, which at low
density is indeed an extremely small number, cf.~Eq.~(\ref{e:ds}).
Therefore, for very small asymmetries already, pairing generated by
the direct interaction between different species becomes impossible.

For larger asymmetry only $p$-wave pairing between like species can
take place, which in leading order in density is given by the
polarization diagram shown in Fig.~\ref{f:dia}(b).  We discuss in the
following the gap of species 1 mediated by the polarization
interaction due to species 2.  Clearly the results are invariant
interchanging 1 and 2.  Quantitatively the relevant interaction kernel
reads at low density \cite{gmb,pet,spp,kag88,bara} \be \lla \kvp
\left| \Gamma_1 \right| \kv \rra = {\Pi_2(|\kv'-\kv|)\over 2} T_0^2
\:,
\label{e:wtot}
\ee with the static Lindhard function (pertaining to the species 2)
\be \Pi_2(q) = -{m k_2 \over \pi^2} \left[{1\over2} + {1-x^2\over
    4x}\ln\left|{1+x\over 1-x}\right| \right] \ ,\quad x = {q\over
  2k_2} \:.
\label{e:pi}
\ee The factor 1/2 in Eq.~(\ref{e:wtot}) corrects for the fact that
conventionally the Lindhard function contains a factor two for the
spin orientations, which is not present in our case.  It should be
noted that it is the absence of exchange diagrams that renders this
low-density polarization interaction attractive in contrast to the
case of one species, where the polarization effects reduce the
$s$-wave BCS gap by a factor $(4e)^{1/3}$ in the low-density limit
\cite{gmb,pet,spp}, see Eq.~(\ref{e:ds}).  It is also remarkable that
the polarization interaction is always attractive,
depending only on the square of the scattering length $a$ \cite{fay}.

Projecting out the $L=1$ partial-wave interaction, one obtains in
particular \cite{kag89,efre} \bea \Gamma_1(k_1,k_1) &=&
{T_0^2 \over 2}\, {1\over 2}\!\int_{-1}^{+1} \!dz\, z\,
\Pi_2(\sqrt{2(1-z)}k_1) \quad,\quad z = \bbox{\widehat{k}}' \cdot
\bbox{\widehat{k}}
\\
&=& -{8 a^2 k_2\over m} {2\ln 2 -1 \over 5} g\Big({k_1\over
  k_2}\Big)\:,
\label{e:linkk}
\eea with \be g(y) = { -1 \over 6(2\ln 2 -1) y^4} \left[ (4-10y^2)
  \ln\left| 1-y^2 \right| - (5+y^2)y^3 \ln\left|{1+y\over 1-y}\right|
  + 4y^2+2y^4 \right] \:.
\label{e:g}
\ee This function is normalized in symmetric matter, $g(1)=1$, and
plotted in Fig.~\ref{f:gap}(a).  The numerical factor $(2\ln 2 -1)/5$
in Eq.~(\ref{e:linkk}) is often replaced by the approximate value
1/13.  Using now the general result for the (angle-averaged) $L$-wave
pairing gap \cite{fay,ander,kohn}, \be \de_L(k_1) \ra c_L\, \mu_1
\exp\!\left[ {2\pi^2 \over m k_1 T_L(k_1,k_1)} \right] \:, \ee and
considering that for a pure polarization interaction $\Gamma_L$, the
leading order in density is $T_L = \Gamma_L$, one obtains \be
\de_1(k_1) = c_1 {k_1^2\over 2m} \exp\!\left[ -{13(\pi/2)^2 \over a^2
    k_1 k_2 g(k_1/k_2)} \right] \:.
\label{e:dtwo}
\ee Taking into account the dependence of this expression on the two
Fermi momenta $k_1$ and $k_2$, the final result for the variation of
the pairing gap with asymmetry $\al$ for fixed total density $\rho$ is
therefore \be {\de(\rho,\al)\over \de(\rho,0)} = (1+\al)^{2/3}
\exp\left[ u(\rho) h(\al) \right]
\label{e:asy}
\ee with \be h(\al) = 1 - { 1 \over (1-\al^2)^{1/3} g\left[
    ((1+\al)/(1-\al))^{1/3} \right] }
\label{e:f}
\ee and the density parameter \be u(\rho) = 13 \left(\pi \over 2{k_F}
  a \right)^2 , \quad {k}_F \equiv (3\pi^2\rho)^{1/3} \:.  \ee The
function $h(\al)$ is displayed in Fig.~\ref{f:gap}(b), where $\al=1$
corresponds to pure 1-matter.
One notes a maximum at $\al \approx 0.478$
with an expansion $h(\al) \approx 0.465-1.343(\al-0.478)^2$.  This
means that the gap at this asymmetry is enhanced by a factor
$e^{0.46u} \approx 10^{0.2u}$ compared to the symmetric case
[$h(0)=0$].  Evidently, in the low-density limit $u\ra\infty$ this
represents an enormous amplification at finite asymmetry.  Around this
peak, the variation of the gap with asymmetry is well described by a
Gaussian with width $\sigma=1/(1.64\sqrt{u})$.  As an illustration,
Fig.~\ref{f:gap}(c) shows the ratio, Eq.~(\ref{e:asy}), for a value of
the density parameter $u = 100$.  A peak of the order of $10^{20}$ is
observed, that becomes rapidly more pronounced and narrower with
decreasing density $\rho$ (increasing $u$), although of course at the
same time the absolute magnitude of the gap decreases strongly with
decreasing density.


Let us now briefly discuss the next-to-leading-order effects, namely
additional contributions of order $(k_i a)^3$ in the denominator of
the exponent of Eq.~(\ref{e:dtwo}).  There are two principal sources
of such effects, which are shown diagramatically in Fig.~\ref{f:pol}.
The first one, Fig.~\ref{f:pol}(a), is the direct $p$-wave interaction
\cite{you} between like species that we have neglected before.
Parametrizing the low-density $p$-wave $T$-matrix in the standard form
$T_1(k,k') \approx 4\pi a_1^3 k k'/ m$, where $a_1$ is the $p$-wave
scattering length, leads to a BCS gap \cite{you} \be {\de \over\mu} =
{8 \over e^{8/3}} \exp\!\left[ {\pi \over 2(k_F a_1)^3}\right] \:.
\ee
The second type of third-order contributions are polarization effects
involving the $s$-wave scattering length $a$.  In contrast to the case
of a one-component system, where there are several relevant diagrams,
in the present two-component system only one diagram exists.  It is
drawn in Fig.~\ref{f:pol}(b).  Unfortunately it can only be computed
numerically, which will not be attempted here.

Finally, to fourth order, there is a large number of diagrams
contributing to the interaction kernel.
Apart from that, at this order it is also necessary to take into
account retardation effects, i.e., the energy dependences of gap
equation, interaction kernel, and self-energy need to be considered
\cite{alex}.
All this can only be done numerically and was performed in
Ref.~\cite{efre} for the case of a one-component system.

In any case, the existence of higher-order corrections will not alter
the main conclusions drawn so far, namely the presence of a strongly
peaked Gaussian variation of the gap with asymmetry.  They will,
however, shift this peak to a different density-dependent location,
and also modify the absolute size of the gap.  Apart from that, the
perturbative approach that is followed here, is clearly limited to the
low-density range $k_F |a| < 1$.  For larger density, different
theoretical methods have to be used \cite{bb}, which is still a
difficult field of current investigation.


In conclusion, we have studied the possibility of pairing in
asymmetric fermion matter composed of two distinct species.
In the low-density limit, the direct $s$-wave interaction between
different species produces a gap $\sim \exp\!\left[{\pi/ 2k_F
    a}\right]$ only for very small asymmetries of the order of
$\de_0/\mu$, Eq.~(\ref{e:sgap}), whereas for larger asymmetries the
polarization-induced $p$-wave attraction between two like species
produces a much smaller gap $\sim \exp\!\left[-13({\pi/ 2k_F
    a})^2\right]$, which extends however in principle over the whole
range of asymmetry.  In practice a sharp maximum at $\al\approx0.478$
($\rho_1/\rho_2 \approx 2.83$, $k_1/k_2\approx$ 1.41) appears.
Explicit expressions for the variation of these gaps with asymmetry
were given.  Higher-order corrections will only quantitatively modify
these particular features of pairing in the low-density regime.
Clearly the experimental observation of both types of pairing is
supposedly difficult, in the first case due to the nearly perfect
symmetry that is required, in the second one due to the extremely
small size of the resulting gap.

\medskip We acknowledge useful discussions with U. Lombardo, A. Ramos,
and P. Schuck.  This work was supported in part by the programs
``Estancias de cient\'{\i}ficos y tecn\'ologos extranjeros en
Espa\~na,'' SGR98-11 (Generalitat de Catalunya), and DGICYT (Spain)
No.~PB98-1247.  J. M. P. wishes to acknowledge support from a
postgraduate fellowship of the Fundaci\'o Universit\`aria Agust\'{\i}
Pedro i Pons.



\begin{figure}
\caption{
  The two possible lowest-order pairing interactions: (a) Direct
  $s$-wave interaction between different species.  (b)
  Polarization-induced $p$-wave interaction between like species.
  $V_0$ and $T_0$ are the $s$-wave ($L=0$) bare potential and
  $T$-matrix between species 1 and 2, respectively.}
\label{f:dia}
\end{figure}

\begin{figure}
\caption{
  BCS momentum distributions of the species 1 and 2 in asymmetric
  superfluid matter.}
\label{f:n}
\end{figure}

\begin{figure}
\caption{
  (a,b) The functions $g$ and $h$, appearing in Eqs.~(\ref{e:g}) and
  (\ref{e:f}), respectively.  (c) The variation of the gap with
  asymmetry, Eq.~(\ref{e:asy}), for a density parameter $u=100$.
  }
\label{f:gap}
\end{figure}

\begin{figure}
\caption{
  Third-order diagrams: (a) Direct $p$-wave interaction, (b)
  Polarization contribution.}
\label{f:pol}
\end{figure}





\newpage
\begin{center}
\begin{figure}
  \includegraphics[totalheight=8.cm,angle=0,bb=400 680 -50
  820]{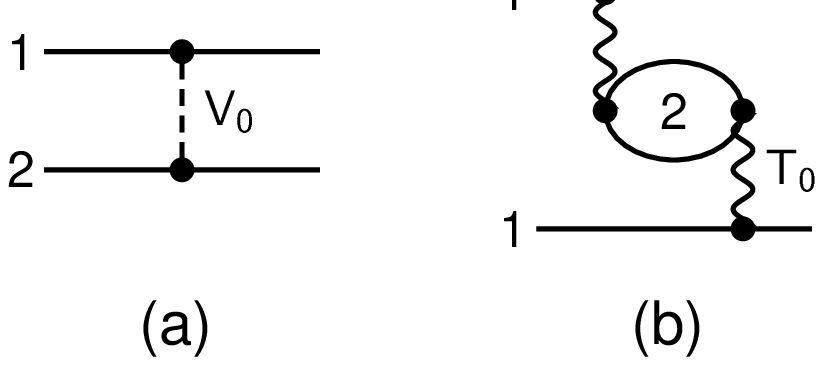}
\end{figure}
\vskip12cm
FIG. 1
\end{center}

\newpage
\begin{center}
\begin{figure}
  \includegraphics[totalheight=8.cm,angle=0,bb=650 520 -80
  750]{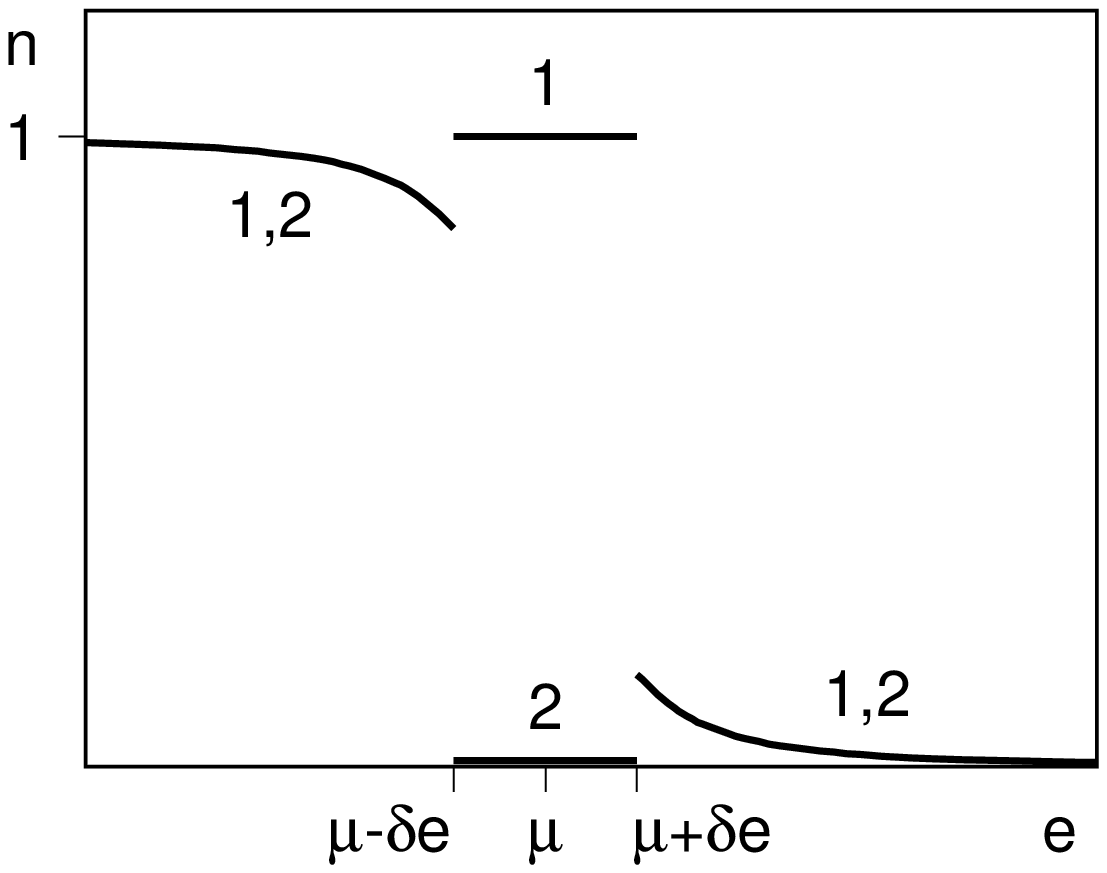}
\end{figure}
\vskip12cm
FIG. 2
\end{center}

\newpage
\begin{center}
\begin{figure}
  \includegraphics[totalheight=4.5cm,angle=90,bb=550 710 70
  820]{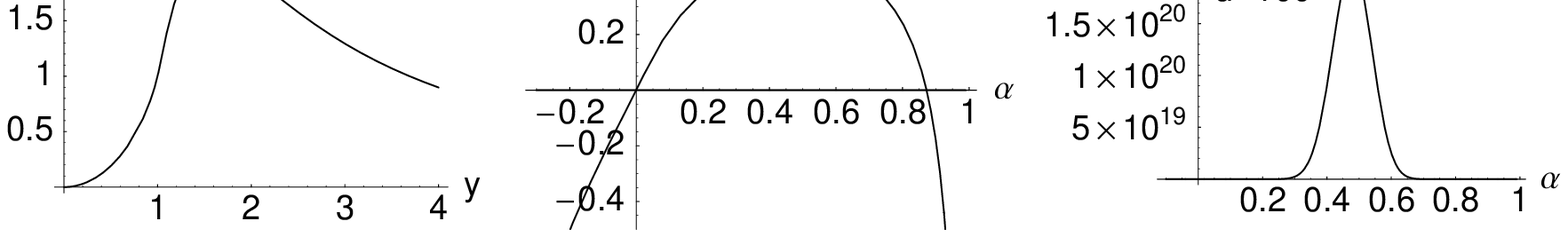}
\end{figure}
\vskip21cm
FIG. 3
\end{center}

\newpage
\begin{center}
\begin{figure}
  \includegraphics[totalheight=8cm,angle=0,bb=400 680 -70
  820]{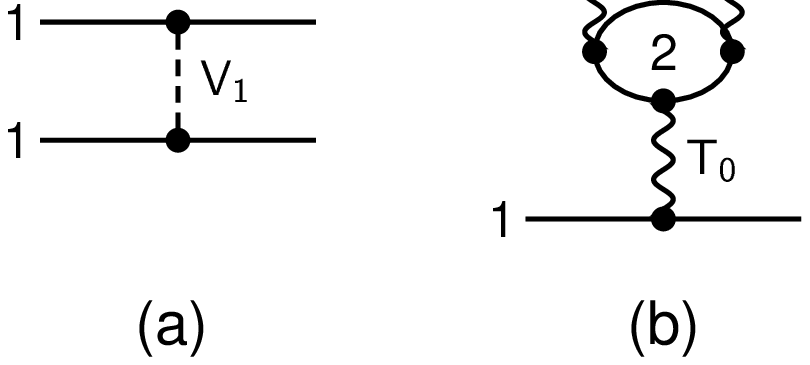}
\end{figure}
\vskip12cm
FIG. 4
\end{center}


\begin{thebibliography}{99}

  
\bibitem{exp} B. De Marco and D. S. Jin, Science {\bf 285}, 1703
  (1999).  \bibitem{gmb} L. P. Gorkov and T. K. Melik-Barkhudarov,
  Sov. Phys. JETP {\bf 13}, 1018 (1961).
\bibitem{pet} H. Heiselberg, C. J. Pethick, H. Smith, and L. Viverit,
  Phys. Rev. Lett. {\bf 85}, 2418 (2000).  \bibitem{spp} H.-J.
  Schulze, A. Polls, A. Ramos, Phys. Rev. {\bf C63}, 044310 (2001).
\bibitem{stoof} H. T. C. Stoof, M. Houbiers, C. A. Sackett, and R. G.
  Hulet,
  Phys. Rev. Lett. {\bf 76}, 10 (1996); \\
  M. Houbiers, R. Ferwerda, H. T. C. Stoof, W. I. McAlexander, C. A.
  Sackett, and R. G. Hulet, Phys. Rev. {\bf A56}, 4864 (1997).
\bibitem{umb} A. Sedrakian, T. Alm, and U. Lombardo,
  Phys. Rev. {\bf C55}, 582 (1997); \\
  A. Sedrakian and U. Lombardo, Phys. Rev. Lett. {\bf 84}, 602 (2000).
\bibitem{akhi} A. I. Akhiezer, A. A. Isayev, S. V. Peletminsky, and A.
  A. Yatsenko, Phys. Rev. {\bf C63}, 021304 (2001).  \bibitem{kag88}
  M. Yu. Kagan and A. V. Chubukov, JETP Lett. {\bf 47}, 614 (1988).
\bibitem{bara} M. A. Baranov, Yu. Kagan, and M. Yu. Kagan, JETP Lett.
  {\bf 64}, 301 (1996).  \bibitem{fay} D. Fay and A. Layzer, Phys.
  Rev. Lett. {\bf 20}, 187 (1968).  \bibitem{kag89} M. Yu. Kagan and
  A. V. Chubukov, JETP Lett. {\bf 50}, 517 (1989).  \bibitem{efre} D.
  V. Efremov, M. S. Marenko, M. A. Baranov, and M. Yu. Kagan, Sov.
  Phys. JETP {\bf 90}, 861 (2000).
\bibitem{ander} P. W. Anderson and P. Morel, Phys. Rev. {\bf 123},
  1911 (1961).  \bibitem{kohn} W. Kohn and J. M. Luttinger, Phys. Rev.
  Lett. {\bf 15}, 524 (1965).  \bibitem{you} L. You and M. Marinescu,
  Phys. Rev. {\bf A60}, 2324 (1999).  \bibitem{alex} A. S. Alexandrov
  and A. A. Golubov, Phys. Rev. {\bf B45}, 4769 (1992).  \bibitem{bb}
  S. Babu and G. E. Brown,
  Ann. Phys. {\bf 78}, 1 (1973); \\
  T. L. Ainsworth and K. S. Bedell, Phys. Rev. {\bf B35}, 8425 (1987).

\end{thebibliography}
\end{document}